# Post-Volcanic Aerosol Altitude and Particle Size Measurements Basing on Twilight Sky Polarimetry

Ugolnikov O.S., Maslov I.A.

Space Research Institute, Russian Academy of Sciences,
84/32 Profsoyuznaya St., Moscow, 117997, Russia

E-mail: ougolnikov@gmail.com, imaslov@iki.rssi.ru

Measurements of the background intensity and polarization of the twilight sky were conducted during the "purple lights" epoch that was potentially caused by the Raikoke volcano eruption in the summer of 2019. An increase in sky brightness paired with a decrease in polarization was registered, especially noticeable in the dusk segment. Using techniques developed in previous works, altitude distribution and the mean particle size of additional aerosol was found. The mean radius of (0.11 μm) is typical for background or moderate eruption conditions, however, aerosols were mostly observed in the upper troposphere and near the tropopause rather than in the stratosphere. A comparison with existing aerosol data after mid-latitude eruptions like the Kasatochi volcano eruption one decade ago showed similar properties, which can be used as confirmation that the "purple lights" event in 2019 was of volcanic nature.

**Keywords:** volcanic aerosol, particle size, scattering, polarization.

## 1. Introduction

Stratospheric aerosols have been the focus of experimental and theoretical studies since their discovery by Junge *et al.* (1961). Appearing in pressure and temperature conditions inconsistent with pure solid or liquid water, they have another chemical nature which was found in the balloon experiment conducted by Rosen (1971). Sulfuric acid droplets can be produced from volcanically originated $SO_2$ if it reaches the stratosphere; this chemical process was described by Weisenschtein *et al.* (1997). It is the explanation for the increase in both density and particle size of stratospheric aerosols after major volcanic eruptions like El Chichon in 1982 and Mt. Pinatubo in 1991 (Deshler *et al.*, 2003).

Solar light scattering by aerosol particles leads to an optical phenomenon that can be observed during twilight when the stratospheric aerosol layer is still being emitted by the Sun, while the underlying tropospheric layer is not. The "purple light" effect was noticed after the eruption of Krakatoa in 1883 (Clark, 1883), long before the direct discovery of stratospheric aerosol particles. Light scattering in particles is described by Mie theory, there is not an excess in the red part of the spectrum. The purple color of the dusk segment is a result of the Chappuis absorption of tangent solar emission by stratospheric ozone (the green and yellow bands) and Rayleigh extinction (mostly in the blue band).

However, even in volcanically-quiescent epochs, the amount of stratospheric aerosol does not completely vanish. Background aerosol particles were studied directly in balloon experiments (Deshler *et al.*, 2006). Having the same chemical composition, these particles appear at an altitude of about 20 km and have a mean radius of around 0.1 μm, while post-volcanic aerosols are observed a little lower (except for the case of strong eruptions) and have a larger particle size (Bauman *et al.*, 2003). The "twilight purple light" event was also observed during volcanically-quiescent periods (Lee & Hernández-Andrés, 2003).

The origin of non-volcanic background sulfate aerosol particles can be carbonyl sulfide OCS (Crutzen, 1976). Having a higher oxidation energy threshold, OCS requires a stronger ultraviolet flux to form the sulfate particles; this is the reason for the altitude difference of background and moderate eruption-originated aerosols (Andersson *et al.*, 2015). Multi-decade balloon measurements (Deshler *et al.*, 2006) did not show any trend of density and properties of the background aerosol; however, the optical depth of aerosols started to rise during the post-Pinatubo epoch in the early 2000s (Solomon *et al.*, 2011). This effect was considered to be possibly anthropogenic (Campbell *et al*., 2015) since the OCS concentration in the atmosphere was higher than in the pre-industrial epoch (Aydin *et al*., 2014). However, this could also be related to moderate eruptions in the 2000s (Neely *et al*., 2013). Tropical volcanoes (Tavurvur in 2006, Nabro in 2011) are able to cause a global increase in stratospheric optical depth due to Brewer-Dobson circulation. In the case of Nabro, $SO_2$ was able to reach the stratosphere through Asian monsoon transport (Bourassa *et al*., 2012). Northern hemisphere volcanoes (Kasatochi in 2008, Sarychev in 2009, Eyjafjallajökull in 2010) had a significant effect in high latitudes. Since the tropopause can be as low as 11-12 km there, a moderate eruption can add a large amount of $SO_2$ into the stratosphere.

The question of the nature of stratospheric aerosol in the early 2000s was to be answered during the following decade, that turned out to be more volcanically-quiescent. A decrease in the optical depth of the aerosol became noticeable in the early 2010s (Ridley *et al*., 2014; Kremser *et al*., 2016). Polarization measurements of the twilight sky (Ugolnikov & Maslov, 2019a) showed the negative trend of this value over eight years, confirming the volcanic nature of stratospheric aerosol changes in the 2000s.

The eruption of the Raikoke volcano (48.3°N, 153.3°E) on the Kuril Islands, off the coast of Russia on June 21st, 2019 had the strongest stratospheric effect in the Northern hemisphere since Nabro in 2011. According to Global Volcanism Program data (Global Volcanism Program, 2019), the plume at its peak reached an altitude close to 13 km. The tropopause was about 12 km above the volcano, according to EOS Aura/MLS satellite data (EOS Team, 2011a). As sulfur dioxide expanded above the Northern Hemisphere, the effects of "purple light" were observed during late summer and early autumn of 2019. The study of the relation of these effects to the Raikoke eruption requires the measurements of the altitude and size of particles and comparison with epochs of previous northern eruptions like Kasatochi or Sarychev. In this paper, this is done through twilight polarization analysis and techniques improved before (Ugolnikov & Maslov, 2019a) for studying the background aerosol.

## 2. Observations

Measurements of twilight sky brightness and polarization were performed by a Wide-Angle Polarization Camera described by Ugolnikov & Maslov (2013) and installed in the vicinity of Moscow, Russia (55.2°N, 37.5°E). The lens field size is about 140°, the landscape restricts the zenith angle of solar vertical points $\zeta$ under consideration to ±50°. Observations are conducted in the spectral band with an effective wavelength of 540 nm and FWHM of about 90 nm. The camera starts working before sunset and terminates after sunrise. The exact position of the camera, field curvature, and atmospheric transparency are fixed by star imaging and photometry during the night.

Bad weather conditions in 2019 restricted the number of clear troposphere twilights at the observation site during the "purple light" epoch in the late summer and early autumn. Here we analyze the moonless morning twilight data of September, 10 and 12, 2019 and compare it with clear-sky moonless morning twilight of autumn 2018.

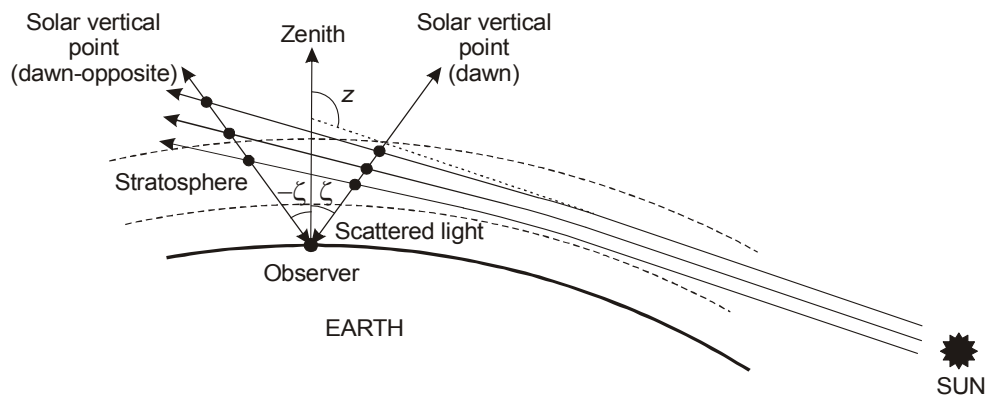

*Figure 1. Single scattering geometry during the twilight.*

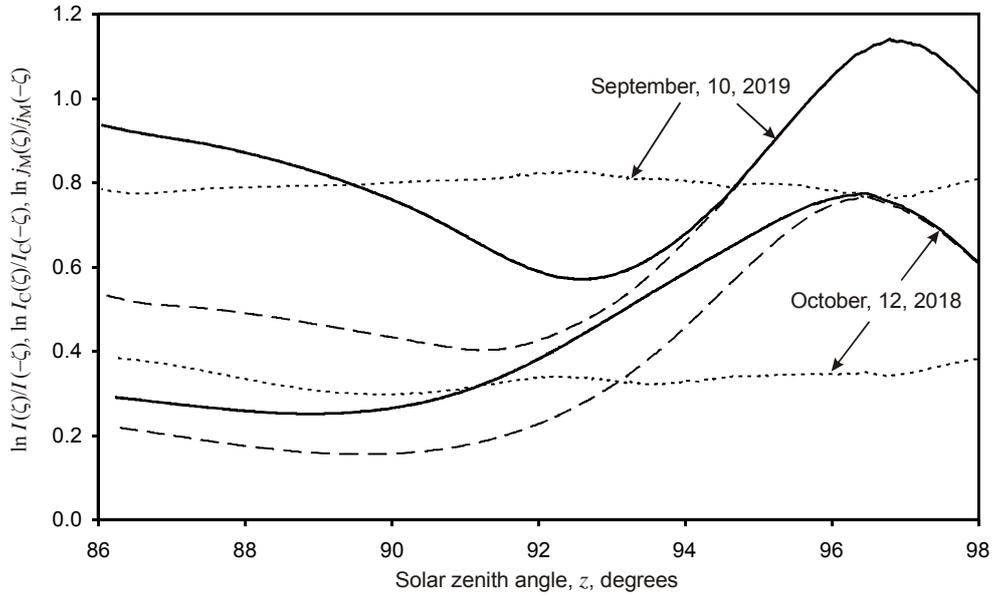

*Figure 2. Logarithm of brightness ratio in symmetrical solar vertical points ($\zeta=45°$) for total background (I, solid lines), without single aerosol scattering ($I_C$, dashed lines), and for multiple scattering ($j_M$, dotted lines) during the twilight of autumn 2018 and 2019.*

Figure 1 shows the geometry of single scattering during the twilight. A significant excess of sky brightness in the dawn segment could appear for two reasons; First, the Earth's shadow is lower in dawn segment, allowing dense lower atmospheric layers to continue being illuminated by the Sun. The shadow altitude difference increases during the dark twilight, and this factor is most applicable to upper atmospheric phenomena like noctilucent clouds (Ugolnikov *et al*., 2016, 2017, Ugolnikov & Maslov, 2019b). Second, according to Mie theory, aerosol particles have an excess of light scattering in the forward direction, multiplying the dawn segment glow. Both effects are clearly seen if the dependence of the brightness ratio in symmetric solar vertical points $I(\zeta)/I(-\zeta)$ is plotted as on the solar zenith angle $z$. This was measured during the clear sky twilight in 2018 and "purple light epoch" in 2019 and shown in Figure 2 (solid lines). While dark twilight asymmetry is seen for both cases (the right side of the figure), the Mie scattering effect appears in the case of increased aerosol scattering (September 2019) on the left side of the figure.

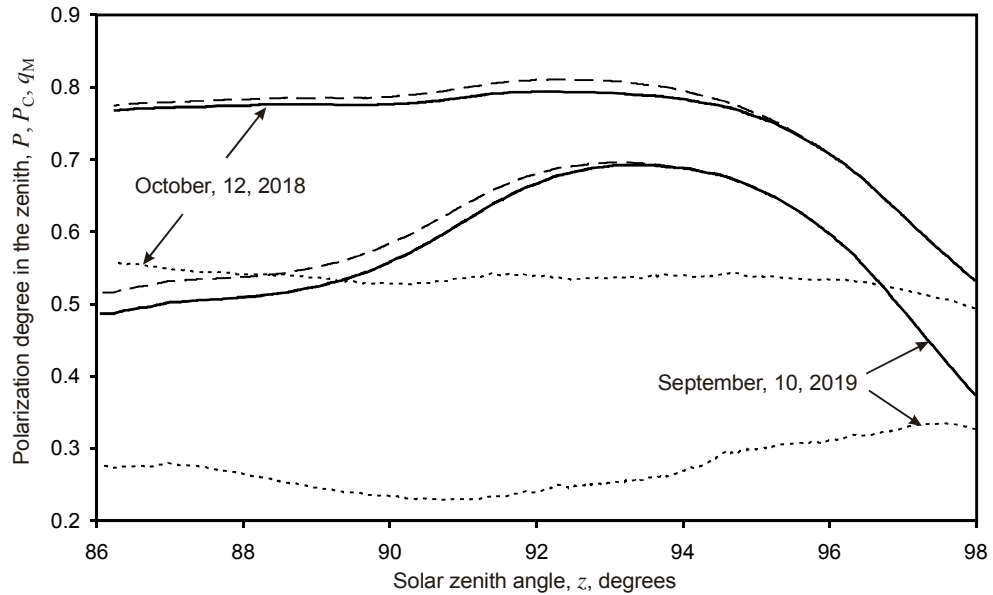

*Figure 3. Degree of polarization of the total sky background (P, solid lines), without single aerosol scattering ($P_C$, dashed lines) and of multiple scattering component ($q_M$, dotted lines) in the zenith during autumn twilights in 2018 and 2019.*

Aerosol effects can be also seen in the dependence of sky polarization on the solar zenith angle shown in Figure 3 for the zenith ($\zeta=0°$). Clear-sky early twilight in the case of the background stratospheric aerosol is characterized by higher zenith polarization reaching 0.8. In autumn 2019, polarization was lower during the whole twilight period including the dark stage, when the troposphere and stratosphere are immersed in the shadow of the Earth. As seen below, it is related to multiple scattering and is also contributed to by volcanic aerosol. The total aerosol effect decreases the zenith polarization near the sunrise to about 0.5.

In this paper, the properties of the aerosol forming the "purple lights" effect will be retrieved using the method described in (Ugolnikov & Maslov, 2019a) and improved in the case of variable particle size. This method is based on the numeric integration of a single scattering component and the empirical properties of multiple scattering. The results can pinpoint whether or not the aerosol actually originated from the Raikoke eruption.

## 3. Retrieval of Aerosol Properties

The procedure of aerosol component retrieval is quite similar to that done in the paper (Ugolnikov & Maslov, 2019a); the process is described there in more detail. Let $I(\zeta, z)$ and $P(\zeta, z)$ be the measured values of intensity and polarization (or first and normalized second Stokes vector components) of the sky background, $z$ – the solar zenith angle, $\zeta$ – the solar vertical point zenith angle which is positive in the dawn area and negative in the opposite part of the vertical (see Figure 1). For the same values of $z$ and $\zeta$, the theoretical values of the brightness of Rayleigh and the aerosol single scattering component: $J_{0R}(\zeta, z, A_i)$ and $J_{0A}(\zeta, z, r, A_i)$. Here the aerosol particle size distribution is assumed to be lognormal with a mean radius $r$ and a distribution width 1.6 (Deshler *et al.*, 2003, Bourassa *et al.*, 2008, Ugolnikov & Maslov, 2018ab). The parameters $A_i$ are equal to the ratio of aerosol and the Rayleigh extinction coefficients at the altitudes $h_i$. These altitudes are defined on a 5-km grid and equal to 5, 10, ... , 60 km. Since the aerosol diminishes tangent solar emission before scattering, Rayleigh scattering brightness $J_{0R}$ also depends on the aerosol characteristics $A_i$. The polarization of molecular and aerosol scattering $p_0(z-\zeta)$ and $p_A(z-\zeta, r)$ is defined by Rayleigh and Mie theory. The single scattering angle is equal to $z-\zeta$, the influence of refraction on this value is not significant (however, refraction is important while calculating the trajectories of solar emission). The refractive index of the sulfate particle is 1.44 (Russell & Hamill, 1984).

Theoretical calculations of single scattering field are performed for each observation day taking into account Rayleigh and aerosol scattering, refraction and Chappuis ozone absorption using the EOS Aura/MLS (EOS Team, 2011ab) profiles of temperature and ozone. The values of Rayleigh and aerosol single scattering intensity measured in observations are equal to:

$$J_R(\zeta, z, A_i) = J_{0R}(\zeta, z, A_i) \cdot (K_1 + K_2\zeta^2);$$
$$J_A(\zeta, z, r, A_i) = J_{0A}(\zeta, z, r, A_i) \cdot (K_1 + K_2\zeta^2). \quad (1)$$

The values of $K_1$ and $K_2$ are initially unknown. This takes into account the camera sensitivity, the flat field and possible uncertainties of atmospheric transparency measured by the images of stars during the night. Subtracting this field from the measured sky background, it is possible to find the intensity and polarization of the multiple scattering field:

$$j_M = I - J_R - J_A;$$
$$q_M = \frac{IP - J_R p_R - J_A p_A}{I - J_R - J_A}. \quad (2)$$

An empirical property of multiple scattering from dark twilight observations (Ugolnikov, 1999; Ugolnikov & Maslov, 2002, 2013) is the similarity of brightness and polarization gradients in symmetric points of solar vertical:

$$\frac{d \ln j_M(\zeta,z)}{dz} = \frac{d \ln j_M(-\zeta,z)}{dz}; \quad \frac{dq_M(\zeta,z)}{dz} = \frac{dq_M(-\zeta,z)}{dz}. \quad (3)$$

Following (Ugolnikov & Maslov, 2019a), this property can be used to run the procedure of the least-squares method and find $r$, $A_i$, $K_1$ and $K_2$ fitting the following criterion:

$$\sum_z \sum_\zeta \left( \frac{d \ln j_M(\zeta,z)}{dz} - \frac{d \ln j_M(-\zeta,z)}{dz} \right)^2 + \left( \frac{dq_M(\zeta,z)}{dz} - \frac{dq_M(-\zeta,z)}{dz} \right)^2 = \min. \quad (4)$$

This is done in a multi-step iteration routine; during each round, $K_1$ and $K_2$ are to be found first based on dark twilight data with a small contribution of single aerosol scattering (96°<$z$<99°) and then all other parameters are to be found. To fix the possibly volcanic aerosol around the tropopause, it is necessary to include the lowest layer corresponding to $A_1$ at 5 km, and the interval 86°<$z$<98° is considered (Ugolnikov & Maslov (2019) chose 10 km and 90°<$z$<98°, respectively).

The model with a constant mean particle radius $r$ is used at the first stage. Table 1 (left) shows the values of $r$ for the autumn twilights of 2018 and 2019. They are close to 0.1 μm, in agreement with the mean radius of particles measured in balloon experiments (Deshler *et al.*, 2003, 2006) and estimated by approximate color and polarization analysis of the twilight sky (Ugolnikov & Maslov, 2018ab) in the background conditions. The close value is taken a priori in OSIRIS satellite limb measurements (Bourassa *et al.*, 2008).

Figure 4 shows the profiles of the ratio of aerosol and Rayleigh extinctions retrieved in the analysis ($A_i$ values). For autumn 2018, a typical Junge layer is seen in the stratosphere. Additional scattering appears in the upper troposphere and near the tropopause in autumn 2019. It is not expanded higher to the stratosphere; the profile of September, 10 is likely to have two separate layers.

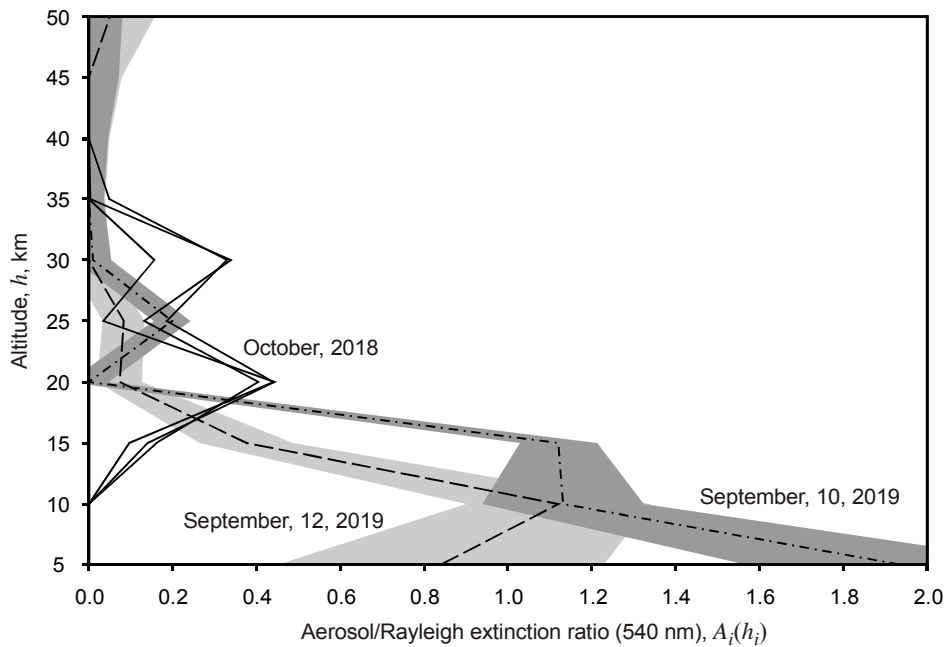

*Figure 4. Vertical profiles of ratio of aerosol and Rayleigh extinction for twilights in 2018 (solid lines) and 2019 (dashed and dashed-dotted lines with error areas).*

| Date (morning twilight) | Mean particle radius (common for all layers) | Mean particle radius at 5-15 km (fixed $r=0.1$ μm above) |
|---|---|---|
| October, 12, 2018 | 0.102 ± 0.011 | — |
| October, 17, 2018 | 0.113 ± 0.023 | — |
| October, 20, 2018 | 0.091 ± 0.019 | — |
| September, 10, 2019 | 0.111 ± 0.012 | 0.110 ± 0.011 |
| September, 12, 2019 | 0.101 ± 0.025 | 0.107 ± 0.024 |

*Table 1. Mean particle radius r for lognormal distribution with width σ=1.6 retrieved as common for all layers and at altitudes 5-15 km with fixed value above.*

To estimate the mean particle size of the aerosol of the lower layer in 2019, it is necessary to run the procedure finding the mean radius *r* at the layers 5-15 km, assuming the fixed value 0.10 μm at 20 km and above. The results are presented in Table 1 (right). The mean radius of the lower aerosol fraction is just a little more (0.11 μm); the difference is below the typical error of *r* estimation. It seems to be congruent with moderate eruption, the same value (0.11 μm) can be found from the polarization analysis after the Tavurvur eruption in 2006 (Ugolnikov & Maslov, 2009).

Figures 2 and 3 show the "clear sky" background characteristics (logarithm of the brightness ratio in symmetrical solar vertical points and polarization in the zenith) after subtraction of single aerosol scattering (dashed lines):

$$\ln \frac{I_C(\zeta, z)}{I_C(-\zeta, z)} = \ln \frac{I(\zeta, z) - J_A(\zeta, z)}{I(-\zeta, z) - J_A(-\zeta, z)};$$

$$P_C(0, z) = \frac{I(0, z) P(0, z) - J_A(0, z) p_A(0, z)}{I(0, z) - J_A(0, z)}. \quad (5)$$

The same characteristics of multiple scattering (ln $j(\zeta, z)/ j(-\zeta, z)$ and $q_M(0, z)$) are shown by dotted lines in the same figures. It can be seen that multiple scattering is characterized by an almost constant brightness ratio in symmetrical solar vertical points and polarization, meeting the conditions (3). In the case of an increased aerosol level in 2019, multiple scattering polarization is significantly less than in the case of the clear atmosphere in 2018, being the primary reason for a decrease in polarization during the light twilight. It shows the principal role of upper tropospheric and stratospheric aerosols in multiple scattering of solar emission during the entire twilight period. It also explains the correlation of sky polarization during the light and dark twilight stages, when single scattering takes place in different atmospheric layers and cannot be strongly correlated (Ugolnikov & Maslov, 2007).

**4. Discussion and conclusion**

In this paper, the authors used twilight polarization analysis to fix additional aerosol scattering in the atmosphere during the "purple light" epoch in the early autumn of 2019 and to find the mean particle size. It is seen that this layer is mostly in the troposphere just slightly expanding to the lower stratosphere. It is separated from the usual stratospheric Junge layer of the background aerosol. This can be explained by the fact that the altitude of the Raikoke eruption plume did not exceed 13 km. Volcanic $SO_2$ is able to form sulfate aerosols at such altitudes, while oxidation of OCS and formation of background particles take place at a higher altitude.

CALIPSO satellite analysis of aerosol extinction after the Kasatochi volcano eruption in 2008 in close location and season (Andersson *et al.*, 2015) showed the downward motion of the layer to the tropopause and upper troposphere in several months after the eruption. It can be also added that the tropopause was about 11 km during the observations in 2019 (EOS Aura/MLS satellite data, EOS Team, 2011a); this is close to the typical altitude of aerosol found here. This is also the altitude of eastward flights of transit airplanes near the place of the observations, sharp shadows of their inversion jets were often observed on the purple glow background. This also rejects the role of post-Raikoke aerosol as condensation nuclei of intensive polar stratospheric clouds in northern latitudes in late 2019 and early 2020. They formed higher in the layer of the deep temperature minimum (<190 K, above 21 km) in a stable polar stratospheric vortex this winter.

The strong dependence of scattered light polarization on the mean particle size described by Mie theory makes the polarization analysis effective for microphysical analysis of aerosol. The mean particle radius is found to be equal to 0.11 μm, this is normal for a weak or moderate eruption. The observational properties of aerosol principally confirm that it originated from the Raikoke eruption in June 2019.